\documentclass[showpacs,showkeys,preprintnumbers]{revtex4}%
\usepackage{amsfonts}
\usepackage{amsmath}
\usepackage{graphicx}
\usepackage{dcolumn}
\usepackage{bm}
\usepackage{amssymb}%
\begin{document}
\title{Is $f_{0}(1710)$ a glueball?}
\author{Stanislaus Janowski, Francesco Giacosa, and Dirk H.\ Rischke}
\affiliation{Institute for Theoretical Physics, Goethe University, Max-von-Laue-Str.\ 1,
D--60438 Frankfurt am Main, Germany }

\begin{abstract}
We study the three-flavor chirally and dilatation invariant extended Linear
Sigma Model with (pseudo)scalar and (axial-)vector mesons as well as a scalar
dilaton field whose excitations are interpreted as a glueball. The model
successfully describes masses and decay widths of quark-antiquark mesons in
the low-energy region up to $1.6$ GeV. Here we study in detail the vacuum
properties of the scalar-isoscalar $J^{PC}=0^{++}$ channel and find that (i) a
narrow glueball is only possible if the vacuum expectation value of the
dilaton field is (at tree-level) quite large (i.e., larger than what lattice
QCD and QCD sum rules suggest) and (ii) that only solutions in which
$f_{0}(1710)$ is predominantly a glueball are found. Moreover, the resonance
$f_{0}(1370)$ turns out to be mainly $(\bar{u}u+\bar{d}d)/\sqrt{2}$ and thus
corresponds to the chiral partner of the pion, while the resonance
$f_{0}(1500)$ is mainly $\bar{s}s.$

\end{abstract}

\maketitle

\section{Introduction}

Glueballs are mesons which are solely made of gluons. The prediction that
glueballs exist dates back to the origin of Quantum Chromodynamics (QCD)
\cite{bag-glueball}: gluons carry color charge and interact strongly with each
other, then it is natural to expect that they form bound states. This
expectation is confirmed by numerous simulations of lattice QCD, see
e.g.\ Refs.\ \cite{mainlattice,Morningstar} and refs.\ therein, in which a
full spectrum of glueballs with different quantum numbers $J^{PC}$ (some of
which are exotic) has been obtained. Although up to now no glueball state has
been unambiguously identified, the search for glueballs will be in the focus
of the future PANDA experiment at FAIR \cite{panda}. The hope is that the
existence of (at least some of the foreseen) glueballs will be ultimately established.

The lightest glueball is predicted by lattice QCD to be a scalar state with a
mass of about $1.6$ GeV \cite{mainlattice,Morningstar}. The search for this
state has been, and still is, in the center of vivid activity in the framework
of low-energy QCD. This state is also important because it is related to two
basic phenomena of QCD: the anomalous breaking of dilatation invariance and
the generation of the gluon condensate. In a widely studied phenomenological
scenario two scalar-isoscalar quarkonia, $\bar{n}n=(\bar{u}u+\bar{d}%
d)/\sqrt{2}$ and $\bar{s}s$, and one bare glueball state mix and form the
scalar resonances $f_{0}(1370),$ $f_{0}(1500)$, and $f_{0}(1710)$
\cite{Close,refs,longglueball,gutsche,ventoreview}. Our aim is to investigate
this system by using a three-flavor chiral effective approach which we
describe in the following.

In Refs.\ \cite{susanna, denisnf2, dilaton, dick, dilnf3, fair13} an effective
model of hadrons, denoted as the extended Linear Sigma Model (eLSM), has been
developed. The mesonic part of the eLSM contains (pseudo)scalar and
(axial-)vector states as well as a scalar dilaton/glueball field and is built
under the requirements of chiral symmetry and dilatation invariance. Chiral
symmetry is broken explicitly (by the current quark masses) and, more
importantly, spontaneously (by the chiral condensate). Dilatation invariance
is explicitly broken by a logarithmic dilaton potential which mimics the trace
anomaly of QCD, according to which gluonic quantum fluctuations give rise to
the fundamental energy scale of QCD, $\Lambda_{QCD}$. The dilaton field, named
$G$, develops a nonzero vacuum expectation value (vev) $G_{0}$ and, in turn,
the fluctuations around the minimum represent the scalar glueball.

In this work we investigate the phenomenology of the scalar glueball in the
eLSM. To this end, we extend both Ref.\ \cite{dilaton} and Ref.\ \cite{dick}.
In Ref.\ \cite{dilaton} the dilaton has been first introduced in the eLSM but
the model has been investigated only for the case of two flavors $N_{f}=2$. In
Ref.\ \cite{dick} a more complete study of the vacuum phenomenology has been
performed in the three-flavor ($N_{f}=3$) version of the eLSM and a good
agreement with experimental data listed in Ref.\ \cite{PDG} for both masses
and decay widths has been achieved. However, the dilaton, although formally
present in order to guarantee dilatation invariant interactions, was not
included when calculating mixing in the scalar-isoscalar sector and the
corresponding decays. In the present paper we close this gap: for $N_{f}=3$
the scalar field $G$ naturally couples to nonstrange and strange mesonic
fields and, in particular, mixes with two scalar-isoscalar quarkonia states.

There are two important and quite general aspects of the physics of the scalar
glueball, which need to be discussed separately.

\textit{1. Is the scalar glueball broad or narrow? }This question is extremely
important for the phenomenology and the assignment of the scalar glueball to
an existing resonance. Yet, conflicting arguments exist: (i) In the
large-$N_{c}$ limit the glueball is predicted to be \emph{narrow}. Namely, the
decay of a bare glueball into two quarkonia (e.g., $G\rightarrow\pi\pi$)
scales as $N_{c}^{-2}$ (for comparison, the decay of a quark-antiquark state
into two quark-antiquark states scales as $N_{c}^{-1}$). Since the
large-$N_{c}$ limit is phenomenologically successful, the quite narrow
resonances $f_{0}(1500)$ and $f_{0}(1710)$ are prime candidates for glueball
states. (ii) In Ref.\ \cite{ellis} it is shown that the decay $G\rightarrow
\pi\pi$ depends on the vev $G_{0}$ of the dilaton field as $G_{0}^{-2}$. The
values of $G_{0}$ can be related to the gluon condensate of QCD by assuming
that the trace anomaly is saturated by the dilaton field. Using the values of
the gluon condensate from either QCD sum rules or lattice QCD calculations, it
turns out that the width of the decay $G\rightarrow\pi\pi$ is \emph{very}
large ($\gtrsim500$ MeV). The authors of Ref.\ \cite{ellis} conclude that the
search for the scalar glueball may be very difficult (if not impossible) if
this state is too broad. [Note that a wide glueball was also discussed in
Refs.\ \cite{klemptwide,ochs,narison}].

In Fig.\ 1 we anticipate our result for the decay of a (bare, i.e.\ unmixed)
scalar glueball into two pions as function of the vev $G_{0}$: for values of
$G_{0}$ which belong to the range obtained by QCD sum rules and lattice QCD
(the vertical band), $G\rightarrow\pi\pi$ is also \emph{very} large, in
complete agreement with Ref.\ \cite{ellis}. The two curves correspond to the
cases with and without (axial-)vector states. One can see that the inclusion
of (axial-)vector degrees of freedom reduces the decay width, but this effect
is not sufficient to make it small enough (when $G_{0}$ is inside the vertical
band). When mixing is taken into account, due to interference phenomena the
strong coupling of $G$ to pions may be reduced for the physical resonances.
Yet, since the quarkonium state $\bar{n}n$ is also expected to be broad, it is
not possible to obtain two narrow resonances $f_{0}(1500)$ and $f_{0}(1710)$
in a three-body mixing scenario. Thus, we realize that we cannot obtain a good
description of the phenomenology of the states $f_{0}(1370),$ $f_{0}(1500),$
and $f_{0}(1710)$ if we impose that $G_{0}$ corresponds to the range given by
QCD sum rules or lattice QCD.
\begin{figure}
[bh]
\begin{center}
\includegraphics[
trim=0.000000in 0.000000in -0.000361in 0.000000in,
height=2.0185in,
width=3.659in
]%
{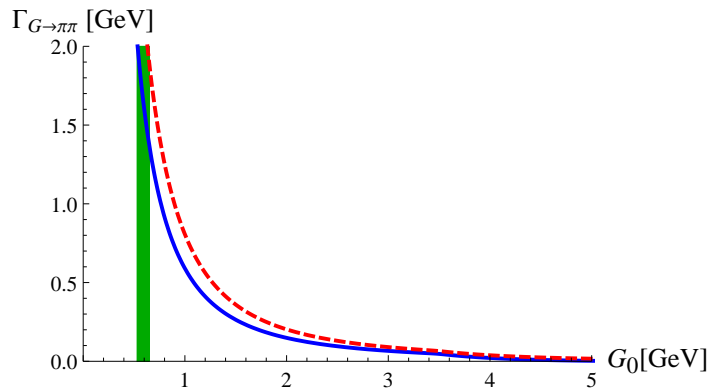}%
\caption{{\small Decay of the pure glueball field into }$\pi\pi${\small \ for
a bare glueball mass }$m_{G}=1525$ MeV. {\small Dashed (online: red) line:
(Axial-)vector mesons are decoupled (}${\small Z}_{\pi}$ ${\small =1}%
${\small ). Solid (online: blue) line: (Axial-)vector mesons are included
(}${\small Z}_{\pi}\neq1$).}%
\label{glu}%
\end{center}
\end{figure}

\textit{2. Assuming that the scalar glueball is narrow, is }$f_{0}%
(1500)$\textit{ or }$f_{0}(1710)$\textit{ mostly gluonic? }A consensus has
grown that the light scalar mesons $f_{0}(500),$ $f_{0}(980),$ $a_{0}(980),$
$K_{0}^{\ast}(800)$ are \emph{not} quark-antiquark states. The possible
assignments are tetraquark or molecular states \cite{tetraquark, lowscalars}.
As a consequence, the scalar quark-antiquark states are located above 1 GeV:
$a_{0}(1450)$ and $K_{0}^{\ast}(1430)$ represent the isovector and isodoublet
$\bar{q}q$ states with $J^{PC}=0^{++}.$ This picture has been confirmed in the
framework of the eLSM \cite{susanna, denisnf2, dilaton, dick}. In particular,
in Ref.\ \cite{dick} a fit to a variety of experimental data has shown that
the scalar states lie between 1 and 2 GeV. Then, if the glueball is a narrow
state, the main question is which of the two resonances $f_{0}(1500)$\textit{
}and\textit{ }$f_{0}(1710)$ contains the largest gluonic amount. In our
previous work \cite{dilaton} two solutions were found, one in which
$f_{0}(1500)$ and one in which $f_{0}(1710)$ was predominantly a glueball (the
former case was slightly favored). Here, we re-analyze this issue in a full
three-flavor study of the eLSM and, quite remarkably, our outcome is now
unique: we find that $f_{0}(1710)$ is predominantly the gluonic state. This
result is in agreement with the original lattice study of
Ref.\ \cite{weingarten}, with (some of) the phenomenological solutions of
Refs.\ \cite{longglueball,cheng} and, interestingly, with the recent lattice
study of $J/\psi$ decays in Ref.\ \cite{chenlattice}. It should be stressed
that the solution in which $f_{0}(1710)$ is a glueball is obtained only if the
value of $G_{0}$ is quite large ($\gtrsim$ 1 GeV). In turn, if this assignment
is correct, this suggests that either the gluon condensate should be larger
than what was previously believed or the dilaton field is not the only
composite field which is responsible for the trace anomaly. Additional fields
may change the values of the parameters in the dilaton potential and thus help
to reconcile the value of $G_{0}$ with lattice QCD and QCD sum rules.

This paper is organized as follows. In Sec.\ II we present the chiral
Lagrangian of our model: the eLSM with a scalar glueball. In Sec.\ III we
discuss our results for the masses and decay widths as well as the three-body
mixing of the resonances $f_{0}(1370)$, $f_{0}(1500)$, and $f_{0}(1710).$
Finally, in Sec.\ IV we present our conclusions and an outlook for future work.

\section{The Model}

As mentioned in the introduction, the aim of this work is to study the
structure of the three scalar-isoscalar resonances $f_{0}(1370)$,
$f_{0}(1500)$, and $f_{0}(1710)$. To this end we use the chiral Lagrangian of
the eLSM developed in Refs.\ \cite{susanna, denisnf2, dilaton, dick}.

\subsection{The dilaton potential}

An essential feature of the eLSM is dilatation invariance together with its
anomalous breaking, which we briefly discuss in the following. The pure
Yang-Mills (YM) Lagrangian reads:%
\begin{equation}%
\mathcal{L}%
_{YM}=-\frac{1}{4}G_{\mu\nu}^{a}G^{a,\mu\nu}\text{ with }G_{\mu\nu}%
^{a}=\partial_{\mu}A_{\nu}^{a}-\partial_{\nu}A_{\mu}^{a}+gf^{abc} A_{\mu}^{b}
A_{\nu}^{c}\text{ ,} \label{ymlag}%
\end{equation}
where $A_{\mu}^{a}$ is the gluon field with $a=1, \ldots,N_{c}^{2}-1=8$,
$G_{\mu\nu}^{a}$ is the gluon field-strength tensor, and $g$ is the QCD
coupling constant. This Lagrangian is classically invariant under dilatation
transformations $x^{\mu}\rightarrow\lambda^{-1}x^{\mu}$, together with
$A_{\mu}^{a}(x)\rightarrow\lambda A_{\mu}^{a}(\lambda x).$ However, when
quantum fluctuations are included and renormalization is carried out, the
coupling constant becomes $g\rightarrow g(\mu),$ where $g(\mu)$ is the
renormalized running coupling which is a function of the energy scale $\mu.$
As a consequence, the divergence of the dilatation (Noether) current does not
vanish:
\begin{equation}
\partial_{\mu}J_{YM,dil}^{\mu}= T_{YM,\mu}^{\mu}=\frac{\beta(g)}{4g}%
\,G_{\mu\nu}^{a}G^{a,\mu\nu}\neq0\text{ ,} \label{ta}%
\end{equation}
where $T_{\text{YM}}^{\mu\nu}$ is the energy-momentum tensor of the YM
Lagrangian and the $\beta$-function is given by $\beta(g)=\partial
g/\partial\ln\mu.$ At the one-loop level $\beta(g)=-bg^{3}$ with
$b=11N_{c}/(48\pi^{2}).$ This implies $g^{2}(\mu)=\left[  2b\ln(\mu
/\Lambda_{YM})\right]  ^{-1},$ where $\Lambda_{YM}\approx200$ MeV is the YM
scale (dimensional transmutation). The expectation value of the trace anomaly
does not vanish and represents the so-called gluon condensate
\begin{equation}
\left\langle T_{YM,\mu}^{\mu}\right\rangle =-\frac{11N_{c}}{48}\left\langle
\frac{\alpha_{s}}{\pi}\,G_{\mu\nu}^{a}G^{a,\mu\nu}\right\rangle =-\frac
{11N_{c}}{48}C^{4}\text{ ,} \label{gc}%
\end{equation}
where
\begin{equation}
C^{4}\approx(0.3\text{-}0.6\text{ GeV})^{4}\text{ .}%
\end{equation}
The numerical values have been obtained through QCD sum rules (lower range of
the interval) \cite{Sumrules} and lattice QCD simulations (higher range of the
interval) \cite{Lattice,digiacomo}. In particular, in Ref.\ \cite{digiacomo}
the value $C\approx0.61$ GeV has been found.

At the composite level one can build an effective theory of the YM sector of
QCD by introducing a scalar dilaton field $G$ which describes the trace
anomaly. The dilaton Lagrangian reads \cite{schechter, migdal}%
\begin{equation}%
\mathcal{L}%
_{dil}=\frac{1}{2}(\partial_{\mu}G)^{2}-\frac{1}{4}\frac{m_{G}^{2}}%
{\Lambda^{2}}\left(  G^{4}\ln\left\vert \frac{G}{\Lambda}\right\vert
-\frac{G^{4}}{4}\right)  \text{ .} \label{dillag}%
\end{equation}
The minimum $G_{0}$ of the dilaton potential is realized for $G_{0}=\Lambda$.
Upon shifting $G\rightarrow G_{0}+G$, a particle with mass $m_{G}$ emerges,
which is interpreted as the scalar glueball. The numerical value has been
evaluated in lattice QCD and reads $m_{G}^{lat}\approx(1.5 -1.7)$ GeV
\cite{Morningstar}. The logarithmic term of the potential explicitly breaks
the invariance under a dilatation transformation. The divergence of the
corresponding current reads%
\begin{equation}
\partial_{\mu}J_{dil}^{\mu}=T_{dil,\,\mu}^{\;\mu} =-\frac{1}{4}\, \frac
{m_{G}^{2}}{\Lambda^{2}}\, G^{4} \longrightarrow-\frac{1}{4}m_{G}^{2}%
\Lambda^{2}\text{ }, \label{dil}%
\end{equation}
where for the last expression we have set $G$ equal to the minimum $G_{0} =
\Lambda$ of the potential.

If we now require that the dilaton field saturates the trace of the dilatation
current, we equate Eq.\ (\ref{gc}) with Eq.\ (\ref{dil}) and obtain:
\begin{equation}
\Lambda\overset{!}{=}\frac{\sqrt{11}\,C^{2}}{2m_{G}}\text{ .}%
\end{equation}

Using $m_{G}\approx(1.5 - 1.7)$ GeV and $C\approx0.61$ GeV \cite{digiacomo}
implies $\Lambda\approx0.4$ GeV. As already shown in Fig.\ 1, if this equation
would hold, the glueball would be too wide when the coupling to ordinary
quarkonia mesons is switched on. A phenomenology with a narrow glueball is
possible only if $\Lambda\gtrsim1$ GeV, see Sec.\ III and the related discussion.

\subsection{The eLSM Lagrangian}

The Lagrangian of the eLSM is built by requiring global chiral $U(3)_{R}\times
U(3)_{L}$ symmetry, dilatation invariance, as well as the discrete symmetries
charge conjugation $C$, parity $P$, and time reversal $T$:%
\begin{align}%
\mathcal{L}%
&  =%
\mathcal{L}%
_{dil}+\text{\textrm{Tr}}[(D^{\mu}\Phi)^{\dag}(D_{\mu}\Phi)]-\text{\textrm{Tr}%
}\left\{  \left[  m_{0}^{2}\left(  \frac{G}{G_{0}}\right)  ^{2}+E\right]
\Phi^{\dag}\Phi\right\}  -\lambda_{1}\left[  \text{\textrm{Tr}}(\Phi^{\dag
}\Phi)\right]  ^{2}-\lambda_{2}\text{\textrm{Tr}}[(\Phi^{\dag}\Phi
)^{2}]\nonumber\\
&  +c_{1}(\text{\textrm{det}}\Phi-\text{\textrm{det}}\Phi^{\dag}%
)^{2}+\mathrm{Tr}[H(\Phi^{\dag}+\Phi)]+\text{\textrm{Tr}}\left\{  \left[
\frac{m_{1}^{2}}{2}\left(  \frac{G}{G_{0}}\right)  ^{2}+\Delta\right]  \left(
L_{\mu}^{2}+R_{\mu}^{2}\right)  \right\} \nonumber\\
&  -\frac{1}{4}\text{\textrm{Tr}}\left(  L_{\mu\nu}^{2}+R_{\mu\nu}^{2}\right)
+\frac{h_{1}}{2}\text{\textrm{Tr}}(\Phi^{\dag}\Phi)\text{\textrm{Tr}}(L_{\mu
}L^{\mu}+R_{\mu}R^{\mu})+h_{2}\text{\textrm{Tr}}(\Phi^{\dag}L_{\mu}L^{\mu}%
\Phi+\Phi R_{\mu}R^{\mu}\Phi^{\dag})\nonumber\\
&  +2h_{3}\text{\textrm{Tr}}(\Phi R_{\mu}\Phi^{\dag}L^{\mu})+ \ldots,
\label{elsm}%
\end{align}
where $D^{\mu}\Phi=\partial^{\mu}\Phi-ig_{1}(L^{\mu}\Phi-\Phi R^{\mu})$ is the
covariant derivative and%

\begin{equation}
\Phi=\sum_{i=0}^{8}(S_{i}+iP_{i})T_{i}=\frac{1}{\sqrt{2}}\left(
\begin{array}
[c]{ccc}%
\frac{\sigma_{N}+a_{0}^{0}+i(\eta_{N}+\pi^{0})}{\sqrt{2}} & a_{0}^{+}+i\pi^{+}
& K_{0}^{\star+}+iK^{+}\\
a_{0}^{-}+i\pi^{-} & \frac{\sigma_{N}-a_{0}^{0}+i(\eta_{N}-\pi^{0})}{\sqrt{2}}
& K_{0}^{\star0}+iK^{0}\\
K_{0}^{\star-}+iK^{-} & \bar{K}_{0}^{\star0}+i\bar{K}^{0} & \sigma_{S}%
+i\eta_{S}%
\end{array}
\right)  \text{ } \label{phimat}%
\end{equation}
is the multiplet of the ordinary scalar ($S$) and pseudoscalar ($P$) mesons
including the bare nonstrange $\sigma_{N}\cong\left(  \bar{u}u+\bar
{d}d\right)  /\sqrt{2}$ and strange $\sigma_{S}\cong\bar{s}s$ fields. Under
$U(3)_{R}\times U(3)_{L}$ chiral transformations $\Phi$ transforms as
$\Phi\rightarrow U_{L}\Phi U_{R}^{\dagger}$. The quantities $L^{\mu}%
=\sum_{i=0}^{8}(V_{i}^{\mu}+A_{i}^{\mu})T_{i}$ and $R^{\mu}=\sum_{i=0}%
^{8}(V_{i}^{\mu}-A_{i}^{\mu})T_{i}$ are the left- and the right-handed vector
matrices, which are linear combinations of the vector and axial-vector
multiplets $V^{\mu}$and $A^{\mu}$. Under chiral transformations, $L_{\mu
}\rightarrow U_{L}L_{\mu}U_{L}^{\dagger}$ and $R_{\mu}\rightarrow U_{R}R_{\mu
}U_{R}^{\dagger}$.

The assignment of the quark-antiquark fields of our model to the resonances
listed by the PDG \cite{PDG} is as follows. (i) In the pseudoscalar sector we
assign the fields $\vec{\pi}$ and $K$ to the physical pion isotriplet and the
kaon isodoublets \cite{PDG}. The bare fields $\eta_{N}\cong i(\bar{u}%
\gamma_{5} u+\bar{d}\gamma_{5} d)/\sqrt{2}$ and $\eta_{S}\cong i \bar{s}%
\gamma_{5} s$ are the nonstrange and strange contributions to the physical
states $\eta$ and $\eta^{\prime}(958)$ \cite{PDG}%
\begin{equation}
\eta=\eta_{N}\cos\varphi_{\eta}+\eta_{S}\sin\varphi_{\eta},\text{ }%
\eta^{\prime}=-\eta_{N}\sin\varphi_{\eta}+\eta_{S}\cos\varphi_{\eta}\text{ },
\label{etaetaprime}%
\end{equation}
where $\varphi_{\eta}\approx-44.6^{\circ}$ is the pseudoscalar mixing angle
\cite{dick}. (ii) As shown in the comprehensive study of Ref.\ \cite{dick},
the scalar $\bar{q}q$ states lie above 1 GeV [in turn, the scalar states below
1 GeV should not be interpreted as $\bar{q}q$ states but as tetraquarks and/or
mesonic molecular states, see
Refs.\ \cite{tetraquark,lowscalars,fariborz,tqchiral}]. Hence, in the scalar
sector we assign the field $\vec{a}_{0}$ to the physical isotriplet resonance
$a_{0}(1450)$ and the scalar kaon isodoublet field $K_{0}^{\star}$ to the
resonance $K_{0}^{\ast}(1430)$ \cite{PDG}. The least clear assignment occurs
in the scalar-isoscalar channel because in the region from $1$ to $2$ GeV
there are three resonances which are listed in Ref.\ \cite{PDG}:
$f_{0}(1370),$ $f_{0}(1500)$, and $f_{0}(1710)$. Only two of them can be
interpreted as predominantly $\bar{q}q$ states while the third one is probably
predominantly a glueball state $G$. The determination of the mixing matrix is
carried out later. (iii) The assignment of the (axial-)vector fields of the
model is straightforward and is presented, together with the corresponding
multiplets, in Appendix \ref{AppA}.

Chiral symmetry is spontaneously broken when $m_{0}^{2}<0.$ Dilatation
symmetry is explicitly broken by the logarithmic term in Eq.\ (\ref{dillag}).
The quantity $G_{0}$ is the vev of the $G$ field, which, in the full version
of the model (\ref{elsm}), is (slightly) larger than $\Lambda$ appearing in
Eq.\ (\ref{dillag}). Moreover, both chiral and dilatation transformations are
explicitly broken by the terms which describe the nonzero bare quark masses of
the mesons, which are proportional to $H=\mathrm{diag} \{h_{N},h_{N},h_{S}\},$
$\Delta=\mathrm{diag}\{0,0,\delta_{S}\},$ and $E=\mathrm{diag}\{0,0,\epsilon
_{S}\}.$ Note that the latter term was not included in Ref.\ \cite{dick}
because it represents a next-to-leading order correction in the expansion in
terms of the quark mass. However, due to the fact that the current mass of the
strange quark is not small, this term is important in our study of the
quark-antiquark scalar state $\sigma_{S}$. The axial anomaly is described by
the determinant term which is invariant under $SU(3)_{R}\times SU(3)_{L}$ but
breaks $U_{A}(1)$. This term which breaks dilatation symmetry and originates
also from the gluon dynamics is responsible for the large mass splitting of
$\eta$ and $\eta^{\prime}$. Note that in the chiral limit (in which
$H=\Delta=E=0$) and neglecting the chiral anomaly, the requirement of
dilatation invariance and analyticity in $G$ ensures that only a finite number
of terms is allowed in our chiral Lagrangian (\ref{elsm}).

Finally, the dots in Eq.\ (\ref{elsm}) indicate further terms which do not
affect the calculations of this work and are therefore neglected, and
additional degrees of freedom which can be studied in the framework of the
eLSM, e.g.\ a pseudoscalar glueball $\tilde{G},$ $J^{PC}=0^{-+},$ which
couples to the ordinary scalar and pseudoscalar mesons. The origin of the
corresponding chiral Lagrangian, $%
\mathcal{L}%
_{\tilde{G}}=ic_{\tilde{G}\Phi}\tilde{G}\left(  \text{\textrm{det}}%
\Phi-\text{\textrm{det}}\Phi^{\dag}\right)  $, comes from the axial anomaly in
the pseudoscalar-isoscalar sector, see details and predictions for branching
ratios in Ref.\ \cite{psg}. An extension of the eLSM to four flavors allows to
describe quite successfully charmed meson masses and decays as well
\cite{walaa}.

\subsection{Lagrangian, masses, and mixing matrix of the scalar-isoscalar
fields}

The three scalar-isoscalar fields $\sigma_{N}\cong(\bar{u}u+\bar{d}d)/\sqrt
{2},$ $\sigma_{S}\cong\bar{s}s$, and $G$ are the only fields of the model with
quantum numbers of the vacuum, $J^{PC}=0^{++}$. In order to study the vev's
and the mixing behavior of these fields we set all the other fields of the
chiral Lagrangian\ (\ref{elsm}) to zero and obtain the scalar-isoscalar
Lagrangian%
\begin{align}%
\mathcal{L}%
_{\sigma_{N}\sigma_{S}G}  &  =%
\mathcal{L}%
_{dil}+\frac{1}{2}(\partial_{\mu}\sigma_{N})^{2}+\frac{1}{2}(\partial_{\mu
}\sigma_{S})^{2}-\frac{m_{0}^{2}}{2}\left(  \frac{G}{G_{0}}\right)
^{2}\left(  \sigma_{N}^{2}+\sigma_{S}^{2}\right) \nonumber\\
&  -\lambda_{1}\left(  \frac{\sigma_{N}^{2}}{2}+\frac{\sigma_{S}^{2}}%
{2}\right)  ^{2}-\frac{\lambda_{2}}{4}\left(  \frac{\sigma_{N}^{4}}{2}%
+\sigma_{S}^{4}\right)  +h_{\sigma_{N}}\sigma_{N}+h_{\sigma_{S}}\sigma
_{S}-\frac{1}{2}\epsilon_{S}\sigma_{S}^{2}\text{ .} \label{isolag}%
\end{align}
Now we perform the shifts of the $J^{PC}=0^{++}$ fields by their vev's,
$\sigma_{N}\rightarrow\sigma_{N}+\phi_{N},\sigma_{S}\rightarrow\sigma_{S}%
+\phi_{S}$, and $G\rightarrow G+G_{0}$, in order to obtain their bare masses
and the bilinear mixing terms $\propto\sigma_{N}\sigma_{S},\propto\sigma_{N}%
G$, and $\propto\sigma_{S}G$. The bare masses of the nonstrange and strange
$\bar{q}q$ fields read%
\begin{equation}
m_{\sigma_{N}}^{2}=C_{1}+2\lambda_{1}\phi_{N}^{2}+\frac{3}{2}\lambda_{2}%
\phi_{N}^{2},\quad m_{\sigma_{S}}^{2}=C_{1}+2\lambda_{1}\phi_{S}^{2}%
+3\lambda_{2}\phi_{S}^{2}+\epsilon_{S}\text{ }, \label{sigmas}%
\end{equation}
where
\begin{equation}
C_{1}=m_{0}^{2}+\lambda_{1}\left(  \phi_{N}^{2}+\phi_{S}^{2}\right)
\label{c1}%
\end{equation}
is a constant \cite{dick} (see Tab.\ \ref{Table1}),%
\begin{equation}
\phi_{N}=Z_{\pi}f_{\pi}\text{ },\quad\phi_{S}=\frac{2Z_{K}f_{K}-\phi_{N}%
}{\sqrt{2}} \;, \label{conds}%
\end{equation}
are the condensates of the nonstrange and strange quark-antiquark states,
where $Z_{\pi/K}$ are the wave-function renormalization constants given in
Eq.\ (\ref{z1}) in Appendix \ref{AppA}, and $f_{\pi/K}$ are the vacuum decay
constants. The bare mass of the scalar glueball reads%
\begin{equation}
M_{G}^{2}=\frac{m_{0}^{2}}{G_{0}^{2}}\left(  \phi_{N}^{2}+\phi_{S}^{2}\right)
+\frac{m_{G}^{2}G_{0}^{2}}{\Lambda^{2}}\left(  1+3\ln\left\vert \frac{G_{0}%
}{\Lambda}\right\vert \right)  \text{ }. \label{glumas}%
\end{equation}
Note that the bare glueball mass also depends on the quark condensates
$\phi_{N}$ and $\phi_{S}$, but correctly reduces to $m_{G}$ in the limit
$m_{0}^{2}=0$ (when quarkonia and the glueball decouple). When quarkonia
couple to the glueball, $m_{0}^{2}\neq0$, the vev $G_{0}$ is given by the
equation
\begin{equation}
-\frac{m_{0}^{2}\Lambda^{2}}{m_{G}^{2}}\left(  \phi_{N}^{2}+\phi_{S}%
^{2}\right)  =G_{0}^{4}\ln\left\vert \frac{G_{0}}{\Lambda}\right\vert \text{
}. \label{G0}%
\end{equation}
The contribution to the tree-level potential, which is of second order in the
fields, reads
\begin{equation}
V^{(2)}=\frac{1}{2} \Sigma^{T} M \Sigma\;, \;\; M \equiv\left(
\begin{array}
[c]{ccc}%
m_{\sigma_{N}}^{2} & 2\lambda_{1}\phi_{N}\phi_{S} & 2m_{0}^{2}\phi_{N}%
G_{0}^{-1}\\[0.1cm]%
2\lambda_{1}\phi_{N}\phi_{S} & m_{\sigma_{S}}^{2} & 2m_{0}^{2}\phi_{S}%
G_{0}^{-1}\\[0.1cm]%
2m_{0}^{2}\phi_{N}G_{0}^{-1} & 2m_{0}^{2}\phi_{S}G_{0}^{-1} & M_{G}^{2}%
\end{array}
\right)  \;,\;\; \Sigma\equiv\left(
\begin{array}
[c]{c}%
\sigma_{N}\\
\sigma_{S}\\
G
\end{array}
\right)  \text{ }. \label{isopot}%
\end{equation}
Following the usual diagonalization procedure, an orthogonal matrix $B$ is
introduced such that the matrix $M^{\prime}=BMB^{T}$\ is diagonal. As a
consequence, $B$ links the bare scalar-isoscalar fields to the physical
resonances:
\begin{equation}
\left(
\begin{array}
[c]{c}%
f_{0}(1370)\\
f_{0}(1500)\\
f_{0}(1710)
\end{array}
\right)  \equiv\Sigma^{\prime}= \left(
\begin{array}
[c]{c}%
\sigma_{N}^{\prime}\\
\sigma_{S}^{\prime}\\
G^{\prime}%
\end{array}
\right)  =B\, \Sigma= B \left(
\begin{array}
[c]{c}%
\sigma_{N}\\
\sigma_{S}\\
G
\end{array}
\right)  \text{ }. \label{mixmat}%
\end{equation}

\subsection{Parameters of the model}

In Ref.\ \cite{dick} a global fit was performed, in which $21$ experimental
quantities were fitted to eleven parameters of the eLSM. Due to their
ambiguous status, scalar-isoscalar mesons were not part of the fit, which
allowed to exclude the coupling constants $\lambda_{1}$ and $h_{1}$ from the
fit. Since we are now explicitly interested in the scalar-isoscalar
resonances, these two coupling constants must be considered, which brings the
number of parameters to 13. Furthermore, in the fit of Ref.\ \cite{dick}, the
glueball was considered to be frozen. This approximation is justifiable in the
large-$N_{c}$ limit because the coupling of one scalar glueball to $m$
ordinary mesons scales as $\sim N_{c}^{-m/2}$. In this study the scalar
glueball is present, which introduces two additional parameters $\Lambda$ and
$m_{G}$, so that we have 15 parameters. Moreover, there is an additional mass
term $\propto\epsilon_{S}$ not present in the study of Ref.\ \cite{dick}, and
thus our chiral Lagrangian (\ref{elsm}) contains 16 parameters. However, since
the parameter $g_{2}$ (which is contained in the dots in Eq.\ (\ref{elsm}))
does not play any role in the present study, we can omit it in the following,
bringing the total number of relevant parameters to be fitted to 15: $\Lambda
$, $m_{G}$, $m_{0}$, $m_{1}$, $\lambda_{1}$, $\lambda_{2}$, $h_{1}$, $h_{2}$,
$h_{3}$, $g_{1}$, $c_{1}$, $h_{0N}$, $h_{0S}$, $\delta_{S}$, $\epsilon_{S}$.
For the calculations in this work we use the values of the parameters $C_{1}$,
$C_{2}$, $\lambda_{2}$, $h_{2}$, $h_{3}$, $g_{1}$, $c_{1}$, $h_{0N}$, $h_{0S}%
$, $\delta_{S}$ determined in Ref. \cite{dick} and shown in Tab.\ \ref{Table1}.%

\begin{table}[h] \centering
$%
\begin{tabular}
[c]{|c|c|c|c|}\hline
\textbf{Parameter} & \textbf{Value} & \textbf{Parameter} & \textbf{Value}%
\\\hline
$C_{1}$ & $-0.918\times10^{6}$ MeV$^{2}$ & $C_{2}$ & $0.413\times10^{6}$
MeV$^{2}$\\\hline
$c_{1}$ & $450\cdot10^{-6}$ MeV$^{-2}$ & $\delta_{S}$ & $0.151\times10^{6}%
$MeV$^{2}$\\\hline
$g_{1}$ & $5.84$ & $\lambda_{2}$ & $68.3$\\\hline
$h_{2}$ & $9.88$ & $h_{3}$ & $3.87$\\\hline
$\phi_{N}$ & $164.6$ MeV & $\phi_{S}$ & $126.2$ MeV\\\hline
\end{tabular}
$%
\caption{Values of the parameters from Ref. \cite{dick}.\label{Table1}}%
\end{table}%

We will perform a fit by using the remaining five free parameters entering
into the model: $\Lambda$, $m_{G}$, $\lambda_{1}$, $h_{1}$, $\epsilon_{S}$.

\section{Results and Discussion}

\subsection{Input and results of the $\chi^{2}$ analysis}

Using the $\chi^{2}$ analysis,
\begin{equation}
\chi^{2}\equiv\chi^{2}(x_{i})=\sum_{j=1}^{8}\left(  \frac{Q_{j}^{th}%
(x_{i})-Q_{j}^{ex}}{\Delta Q_{j}^{ex}}\right)  ^{2}\;\;\text{, with
}i=1,\ldots,5\text{ ,}%
\end{equation}
we fit eight experimental quantities to the five parameters $x_{i}=\Lambda$,
$m_{G}$, $\lambda_{1}$, $h_{1}$, $\epsilon_{S}$ of our chiral model summarized
in Tabs.\ \ref{Table2} and \ref{Table3}.

For the mass of $f_{0}(1370)$ we use the value $M_{f_{0}(1370)}=(1350\pm150)$
MeV and we increase the experimental errors of $M_{f_{0}(1500)}=(1505\pm
6)$\ MeV and $M_{f_{0}(1710)}=(1720\pm6)$ \cite{PDG} to $5\%$ of their
physical values. This procedure was also applied in Ref.\ \cite{dick}, arguing
that the precision of our model cannot be better than $5\%$ since it does not
account e.g.\ for isospin breaking effects. Moreover, in order to better
constrain the fit we use the value $\Gamma_{f_{0}(1370)\rightarrow\pi\pi}=325$
MeV \cite{Bugg} together with an estimated uncertainty [not given in
\cite{Bugg}] of about $100$ MeV. The parameters in Tab.\ \ref{Table2}, for
which $\chi^{2}/d.o.f.\approx0.35$ was achieved, and the masses as well as the
decay widths of the scalar-isoscalar resonances in Tab.\ \ref{Table3}
correspond to the solution in which $\sigma_{N}^{\prime}\equiv f_{0}%
(1370)\cong(\bar{u}u+\bar{d}d)/\sqrt{2}$ is predominantly a nonstrange,
$\sigma_{S}^{\prime}\equiv f_{0}(1500)\cong\bar{s}s$ predominantly a strange
$\bar{q}q$ state, and $G^{\prime}\equiv f_{0}(1710)$ predominantly a glueball state.

%

\begin{table}[h] \centering
\begin{tabular}
[c]{|c|c|}\hline
\textbf{Parameter} & \textbf{Value}\\\hline
$\Lambda$ & $3297\,[$MeV$]$\\\hline
$m_{G}$ & $1525\,[$MeV$]$\\\hline
$\lambda_{1}$ & $6.25$\\\hline
$h_{1}$ & $-3.22$\\\hline
$\epsilon_{S}$ & $0.4212 \times10^{6}\,[$MeV$^{2}]$\\\hline
\end{tabular}%
\caption{Parameters obtained from the fit with the solution: \{$\sigma_N'
$, $\sigma_S' $, $G'$\} $\equiv$
\{$f_0(1370)$, $f_0(1500)$, $f_0(1710)$\}.\label{Table2}}%
\end{table}%

\bigskip%

\begin{table}[h] \centering
\begin{tabular}
[c]{|c|c|c|}\hline
\textbf{Quantity} & \textbf{Fit [MeV]} & \textbf{Exp. [MeV]}\\\hline
$M_{f_{0}(1370)}$ & $1444$ & $1200$-$1500$\\\hline
$M_{f_{0}(1500)}$ & $1534$ & $1505\pm6$\\\hline
$M_{f_{0}(1710)}$ & $1750$ & $1720\pm6$\\\hline
$f_{0}(1370)\rightarrow\pi\pi$ & $423.6$ & -\\\hline
$f_{0}(1500)\rightarrow\pi\pi$ & $39.2$ & $38.04\pm4.95$\\\hline
$f_{0}(1500)\rightarrow K\bar{K}$ & $9.1$ & $9.37\pm1.69$\\\hline
$f_{0}(1710)\rightarrow\pi\pi$ & $28.3$ & $29.3\pm6.5$\\\hline
$f_{0}(1710)\rightarrow K\bar{K}$ & $73.4$ & $71.4\pm29.1$\\\hline
\end{tabular}%
\caption{Fit with the solution: \{$\sigma_N'$, $\sigma_S' $, $G'$\} $\equiv
$ \{$f_0(1370)$, $f_0(1500)$, $f_0(1710)$\}. \label{Table3}}%
\end{table}%

The bare fields $\sigma_{N}\cong\left(  \bar{u}u+\bar{d}d\right)  /\sqrt{2},$
$\sigma_{S}\cong\bar{s}s$, and $G$ generate the resonances $f_{0}(1370),$
$f_{0}(1500)$, and $f_{0}(1710)$, where the corresponding mixing matrix $B$,
cf.\ Eq.\ (\ref{mixmat}), is given by
\begin{equation}
B=\left(
\begin{array}
[c]{ccc}%
-0.91 & 0.24 & -0.33\\
0.30 & 0.94 & -0.17\\
-0.27 & 0.26 & 0.93
\end{array}
\right)  \text{ }, \label{mixmat1}%
\end{equation}
which implies the following admixtures of the bare fields to the resonances:
\begin{align}
f_{0}(1370):  &  \quad83\%\,\sigma_{N},\quad6\%\,\sigma_{S},\quad11\%\,G\text{
,}\nonumber\\
f_{0}(1500):  &  \quad9\%\,\sigma_{N},\quad88\%\,\sigma_{S},\quad3\%\,G\text{
,}\label{admix1}\\
f_{0}(1710):  &  \quad8\%\,\sigma_{N},\quad6\%\,\sigma_{S},\quad86\%\,G\text{
.}\nonumber
\end{align}
The parameters $\lambda_{1}$ and $h_{1}$ are small, in agreement with the
large-$N_{c}$ expectation: they scale as $1/N_{c}^{2}$ and not as $1/N_{c}$.
The numerical value $\Lambda\approx3.3$ GeV suppresses the quarkonium-glueball
mixing: this is why the admixtures in Eq.\ (\ref{admix1}) are small.

In the pure YM sector the vev of the dilaton field $G$ is given by
$G_{0}=\Lambda.$ The numerical value $\Lambda\approx3.3$ GeV implies that the
resulting gluon condensate in pure YM, which is parametrized by the constant
$C$ defined in Eq.\ (\ref{gc}), reads $C\approx1.8$ GeV, which is a factor $3$
larger than the lattice value $C\approx0.61$ GeV obtained in
Ref.\ \cite{digiacomo}. When quarks are included, the value of $G_{0}$ is such
that $G_{0}\approx\Lambda$ to a very good level of precision, see
Eq.\ (\ref{G0}). Similarly, using Eq.\ (\ref{glumas}) the value of the bare
glueball mass in the presence of quarks reads $M_{G}\approx m_{G}$. The fact
that $G_{0}\approx\Lambda$ and $M_{G}\approx m_{G}$ is also a consequence of
the large value of $\Lambda.$ (For small $\Lambda\lesssim0.6$ GeV the
differences are larger.)

Our determination of the parameter $C$ is based on the assumption that the
glueball is narrow, see Fig.\ 1 and the discussion in the introduction. If
this assumption does not hold, the glueball is very broad (and would probably
remain undetected). If, however, the narrow-glueball hypothesis is correct,
our results imply that either (i) the value of the constant $C$ cannot be
directly compared to the corresponding one appearing in lattice QCD or QCD sum
rules (which is entirely possible because there may be corrections to the
tree-level Lagrangian (\ref{dillag}) arising from renormalization), or (ii)
that it is \emph{not} allowed to assume that the dilaton field saturates the
trace anomaly. In turn, Eq.\ (\ref{dil}) would not hold and other
contributions should appear in order to reconcile the mismatch.

The stability of the fit has been also tested by repeating the minimum search
for different values of the parameters, by increasing or reducing the errors
in some channels and by including and/or removing some experimental
quantities. The same pattern has always been found: in all solutions the
resonance $f_{0}(1710)$ is (by far) predominantly a glueball, while
$f_{0}(1370)$ and $f_{0}(1500)$ are predominantly $(\bar{u}u+\bar{d}%
d)/\sqrt{2}$ and $\bar{s}s$ quark-antiquark states, respectively.

\subsection{Consequences of the $\chi^{2}$ analysis}

As a consequence of our fit we calculate the decay processes given in
Tab.\ \ref{Table4}. We discuss our results in the following:

(a) At present, the different decay channels of the resonance $f_{0}(1370)$
are experimentally not yet well known because conflicting experimental results
exist \cite{PDG}. Only the full decay width is listed in Ref.\ \cite{PDG}:
$\Gamma_{f_{0}(1370)}^{exp}=(200 - 500)$ MeV. In our solution the dominant
decay channel of $f_{0}(1370)$ is the one into two pions with a decay width of
about $400$ MeV. This corroborates that $f_{0}(1370)$ is predominantly a
nonstrange $\bar{q}q$ state as also found in
Refs.\ \cite{denisnf2,dilaton,dick}. The total decay width of $f_{0}(1370)$
obtained with the parameters of Tab.\ \ref{Table2} is $598$ MeV. In addition,
we found non-negligible contributions from the decays $f_{0}(1370)\rightarrow
\eta\eta$ and $f_{0}(1370)\rightarrow\rho\rho\rightarrow4\pi$ (where in the
latter case we have integrated over the corresponding $\rho$ spectral
function). These results are in qualitative agreement with the experimental
analysis of Ref.\ \cite{Bugg}, where $\Gamma_{f_{0}(1370)\rightarrow\pi\pi
}=325$ MeV, $\Gamma_{f_{0}(1370)\rightarrow4\pi}\approx50$ MeV, and
$\Gamma_{f_{0}(1370)\rightarrow\eta\eta}/\Gamma_{f_{0}(1370)\rightarrow\pi\pi
}=0.19\pm0.07$. Note that the channel $f_{0}(1370)\rightarrow f_{0}%
(500)f_{0}(500)\rightarrow4\pi$ is not included in our model, so our
determination of the $4\pi$--decay mode is not complete.

(b) When omitting the quantity $\Gamma_{f_{0}(1370)\rightarrow\pi\pi}$ from
the fit, a solution with a similar phenomenology is found. However, the state
$f_{0}(1370)$ would be somewhat too wide ($\approx700$ MeV.) This is why we
have decided to include the quantity $\Gamma_{f_{0}(1370)\rightarrow\pi\pi
}=325$ MeV \cite{Bugg} in the fit.

(c) The decay channel $f_{0}(1500)\rightarrow\eta\eta$ turns out to be in good
agreement with the experiment.

(d) Experimentally, there is also a sizable contribution of the channel
$f_{0}(1500)\rightarrow4\pi$: $\Gamma_{f_{0}(1500)\rightarrow4\pi}%
^{exp}=(54.0\pm7.1)$ MeV. We have calculated the decay of $f_{0}(1500)$ into
$4\pi$ only through the intermediate $\rho\rho$ state (as in the case of
$f_{0}(1370)$ and $f_{0}(1710)$, respectively, including the $\rho$ spectral
function). We found that this decay channel is strongly suppressed. However,
we expect a further (and much larger) contribution to this decay channel
through the intermediate state of two $f_{0}(500)$ resonances, but
$f_{0}(500)$ is not implemented in the present model, see outlook I in Sec.\ IV.

(e) The decay channel $f_{0}(1710)\rightarrow\eta\eta$ is slightly larger than
the experiment.

(f) In comparison with the $N_{f}=2$ results of Ref.\ \cite{dilaton}, we now
find that the decay channel $f_{0}(1710)\rightarrow\rho\rho\rightarrow4\pi$ is
strongly suppressed. The reason is the scaling $\Gamma_{f_{0}(1710)\rightarrow
\rho\rho\rightarrow4\pi}\propto$ $1/G_{0}$. This is indeed an important point:
in Ref.\ \cite{dilaton} two scenarios were phenomenologically acceptable, one
in which $f_{0}(1500)$ and one in which $f_{0}(1710)$ is predominantly a
glueball. The latter case was, however, slightly disfavored because
$\Gamma_{f_{0}(1710)\rightarrow\rho\rho\rightarrow4\pi}$ was too large in
virtue of the vev $G_{0}\sim\Lambda$, which was much smaller in that case. A
solution of that type was possible because only one quarkonium existed and
less experimental information was taken into account.

%

\begin{table}[h] \centering
$%
\begin{tabular}
[c]{|c|c|c|}\hline
\textbf{Decay Channel} & \textbf{Our Value [MeV]} & \textbf{Exp.
[MeV]}\\\hline
$f_{0}(1370)\rightarrow K\bar{K}$ & $117.5$ & -\\\hline
$f_{0}(1370)\rightarrow\eta\eta$ & $43.3$ & -\\\hline
$f_{0}(1370)\rightarrow\rho\rho\rightarrow4\pi$ & $13.8$ & -\\\hline
$f_{0}(1500)\rightarrow\eta\eta$ & $4.7$ & $5.56\pm1.34$\\\hline
$f_{0}(1500)\rightarrow\rho\rho\rightarrow4\pi$ & $0.2$ & $>54.0\pm
7.1$\\\hline
$f_{0}(1710)\rightarrow\eta\eta$ & $57.9$ & $34.3\pm17.6$\\\hline
$f_{0}(1710)\rightarrow\rho\rho\rightarrow4\pi$ & $0.5$ & -\\\hline
\end{tabular}
$%
\caption{Consequences of the fit with the solution: \{$\sigma_N'
$, $\sigma_S' $, $G'$\} $\equiv
$ \{$f_0(1370)$, $f_0(1500)$, $f_0(1710)$\}.\label{Table4}}%
\end{table}%

\section{Conclusions and Outlook}

\subsection{Conclusions}

In the present paper, the scalar glueball state of the extended Linear Sigma
Model, which was considered to be frozen in Ref.\ \cite{dick}, was elevated to
a dynamical degree of freedom. We then studied a three-state mixing scenario
in the scalar-isoscalar sector, where a nonstrange and a strange
quark-antiquark state mix with the glueball to produce the physical resonances
$f_{0}(1370)$, $f_{0}(1500)$, and $f_{0}(1710)$. We have found that the
resonance $f_{0}(1710)$ is predominantly a glueball state, as was also
obtained in Refs.\ \cite{longglueball, cheng, weingarten, chenlattice}.
Moreover, we find that the state $f_{0}(1370)$ is predominantly a nonstrange
quarkonium $\left(  \bar{u}u+\bar{d}d\right)  /\sqrt{2}$ and $f_{0}(1500)$ a
strange quarkonium $\bar{s}s$. Our solution implies that the gluon condensate
$G_{0}$ arising from the tree-level dilaton potential (\ref{dillag}) is about
a factor 3 larger than the one obtained in lattice QCD and QCD sum rule
calculations. As already noticed in Ref.\ \cite{ellis}, this is quite natural
if one wants to obtain a narrow glueball state.

\subsection{Outlook}

\subsubsection{Inclusion of light tetraquark fields}

One should include the nonet of light scalar states $f_{0}(500)$,
$f_{0}(980),$ $a_{0}(980)$, and $K_{0}^{\ast}(800)$, which then allows to
describe all scalar states up to $1.7$ GeV. Indeed, in the two-flavor case the
resonance $f_{0}(500)$ as a tetraquark/molecular field has been already
included in a simplified version of the eLSM \cite{heinz}, in which chiral
symmetry restoration at nonzero temperature has been studied, and in the
extension of the eLSM to the baryonic sector \cite{elsmdensity}. The role of
$f_{0}(500)$ is important because it induces a strong attraction between
nucleons and affects the properties of nuclear matter at nonzero density.

In the three-flavor case chiral models with tetraquark fields but without
(axial-)vector mesons were studied \cite{tqchiral,fariborz,huang}. The
isovector resonances $a_{0}(1450)$ and $a_{0}(980)$ arise as a mixing of a
bare quark-antiquark and a bare tetraquark/molecular field configuration. A
similar situation holds in the isodoublet sector for $K_{0}^{\ast}(1430)$ and
$K_{0}^{\ast}(800)$. The mixing angle turns out to be small \cite{tqchiral}.
In the scalar-isoscalar sector one has a mixing of five bare fields, which
leads to the five resonances $f_{0}(500)$, $f_{0}(980)$, $f_{0}(1370)$,
$f_{0}(1500),$ and $f_{0}(1710)$ \cite{fariborz}.

In the framework of the eLSM, the inclusion of the light scalars should also
contain their coupling to (axial-)vector degrees of freedom as well as to the
dilaton field. A variety of decays, such as the decays of the light scalars
($f_{0}(500)\rightarrow\pi\pi,$ $f_{0}(980)\rightarrow KK$, etc.) as well as
decays into them ($a_{1}(1230)\rightarrow f_{0}(500)\pi$, $f_{0}%
(1500)\rightarrow f_{0}(500)f_{0}(500)$, etc.) can be studied. Moreover, the
mixing in the isovector, isodoublet, and -- most importantly -- in the
isoscalar sector can be investigated in such a framework.

\subsubsection{Inclusion of other glueball fields}

In Ref.\ \cite{psg} the pseudoscalar glueball has been coupled to the eLSM and
its branching ratios have been calculated. The mass of the pseudoscalar
glueball is about $2.6$ GeV \cite{mainlattice}, which is already in the reach
of the PANDA experiment \cite{panda}. Lattice QCD predicts a full tower of
heavier gluonic states with various quantum numbers, such as $J^{PC}%
=1^{--},1^{+-},2^{++}, \ldots$ \cite{mainlattice,Morningstar}. These glueball
states can be easily implemented in the eLSM in a chirally invariant way: the
decays can be evaluated, thus giving useful information about the properties
of these (still hypothetical) glueballs. The search for theses states could be
simplified if clear theoretical input about their decay pattern is known.

\section*{Acknowledgments}

The authors thank I.\ Mishustin and D.\ Parganlija for useful discussions.
S.J.\ acknowledges support from H-QM and HGS-HIRe.

\appendix

\section{Details of the extended Linear Sigma Model}

\label{AppA}

\subsection{Vector and (axial-)vector multiplets and renormalization
constants}

The left-handed and right-handed (axial-)vector fields of the eLSM are
contained in the multiplets \cite{dick}
\begin{equation}
L^{\mu}=\sum_{i=0}^{8}(V_{i}^{\mu}+A_{i}^{\mu})T_{i}=\frac{1}{\sqrt{2}}\left(
\begin{array}
[c]{ccc}%
\frac{\omega_{N}^{\mu}+\rho^{\mu0}}{\sqrt{2}}+\frac{f_{1N}^{\mu}+a_{1}^{\mu0}%
}{\sqrt{2}} & \rho^{\mu+}+a_{1}^{\mu+} & K^{\star\mu+}+K_{1}^{\mu+}\\
\rho^{\mu-}+a_{1}^{\mu-} & \frac{\omega_{N}^{\mu}-\rho^{\mu0}}{\sqrt{2}}%
+\frac{f_{1N}^{\mu}-a_{1}^{\mu0}}{\sqrt{2}} & K^{\star\mu0}+K_{1}^{\mu0}\\
K^{\star\mu-}+K_{1}^{\mu-} & \bar{K}^{\star\mu0}+\bar{K}_{1}^{\mu0} &
\omega_{S}^{\mu}+f_{1S}^{\mu}%
\end{array}
\right)  \;, \label{l}%
\end{equation}
and
\begin{equation}
R^{\mu}=\sum_{i=0}^{8}(V_{i}^{\mu}-A_{i}^{\mu})T_{i}=\frac{1}{\sqrt{2}}\left(
\begin{array}
[c]{ccc}%
\frac{\omega_{N}^{\mu}+\rho^{\mu0}}{\sqrt{2}}-\frac{f_{1N}^{\mu}+a_{1}^{\mu0}%
}{\sqrt{2}} & \rho^{\mu+}-a_{1}^{\mu+} & K^{\star\mu+}-K_{1}^{\mu+}\\
\rho^{\mu-}-a_{1}^{\mu-} & \frac{\omega_{N}^{\mu}-\rho^{\mu0}}{\sqrt{2}}%
-\frac{f_{1N}^{\mu}-a_{1}^{\mu0}}{\sqrt{2}} & K^{\star\mu0}-K_{1}^{\mu0}\\
K^{\star\mu-}-K_{1}^{\mu-} & \bar{K}^{\star\mu0}-\bar{K}_{1}^{\mu0} &
\omega_{S}^{\mu}-f_{1S}^{\mu}%
\end{array}
\right)  \;. \label{r}%
\end{equation}
The assignment of the fields in Eq.\ (\ref{l}) and (\ref{r}) to the physical
resonances is as follows. In the $J^{PC}=1^{--}$ sector the nonstrange
$\omega_{N}^{\mu}$ and the strange $\omega_{S}^{\mu}$\ field represent the
resonance $\omega(782)$ and $\phi(1020)$, respectively. The isotriplet field
$\vec{\rho}^{\mu}$ and the isodoublet fields $K^{\star\mu}$ correspond to the
resonance $\rho(770)$ and $K^{\ast}(1410)$, respectively. In the
$J^{PC}=1^{++}$ sector the nonstrange $f_{1N}^{\mu}$ and the strange
$f_{1S}^{\mu}$ field are assigned to the resonance $f_{1}(1285)$ and
$f_{1}(1420)$. The isotriplet field $\vec{a}_{1}^{\mu}$ is identified with the
resonance $a_{1}(1260)$. Finally, the isodoublet fields $K_{1}$ corresponds to
a mixture of $K_{1}(1270)$ and $K_{1}(1400)$, for details see
Ref.\ \cite{florianlisa}.

Spontaneous breaking of chiral symmetry induces bilinear terms in the
Lagrangian of the eLSM which can be eliminated by shifting the (axial-)vector
fields as follows \cite{dick},%
\begin{equation}
f_{1N/S}^{\mu}\rightarrow f_{1N/S}^{\mu}+Z_{\eta_{N/S}}w_{f_{1N/S}}%
\partial^{\mu}\eta_{N/S},\quad a_{1}^{\mu\pm,0}\rightarrow a_{1}^{\mu\pm
,0}+Z_{\pi}w_{a_{1}}\partial^{\mu}\pi^{\pm,0}\;, \label{f1shift}%
\end{equation}%
\begin{equation}
K_{1}^{\mu\pm,0,\bar{0}}\rightarrow K_{1}^{\mu\pm,0,\bar{0}}+Z_{K}w_{K_{1}%
}\partial^{\mu}K^{\pm,0,\bar{0}}\text{ ,}\quad K^{\star\mu\pm,0,\bar{0}%
}\rightarrow K^{\star\mu\pm,0,\bar{0}}+Z_{K^{\star}}w_{K^{\star}}\partial
^{\mu}K_{0}^{\star\pm,0,\bar{0}}\;. \label{k1shift}%
\end{equation}
After performing this procedure additional kinetic terms occur. In order to
remove the latter a redefinition of the (pseudo)scalar fields is required,%
\begin{equation}
\pi^{\pm,0}\rightarrow Z_{\pi}\pi^{\pm,0}\;,\quad\eta_{N/S}\rightarrow
Z_{\eta_{N/S}}\eta_{N/S}\;, \label{psshift}%
\end{equation}%
\begin{equation}
K^{\pm,0,\bar{0}}\rightarrow Z_{K}K^{\pm,0,\bar{0}}\;,\quad K_{0}^{\star
\pm,0,\bar{0}}\rightarrow Z_{K^{\star}}K_{0}^{\star\pm,0,\bar{0}}\;,
\label{sshift}%
\end{equation}
where
\begin{equation}
Z_{\pi}=Z_{\eta_{N}}=\frac{m_{a_{1}}}{\sqrt{m_{a_{1}}^{2}-g_{1}^{2}\phi
_{N}^{2}}}\;,\quad Z_{K}=\frac{2m_{K_{1}}}{\sqrt{4m_{K_{1}}^{2}-g_{1}^{2}%
(\phi_{N}+\sqrt{2}\phi_{S})^{2}}}\;, \label{z1}%
\end{equation}%
\begin{equation}
Z_{K^{\star}}=\frac{2m_{K^{\star}}}{\sqrt{4m_{K^{\star}}^{2}-g_{1}^{2}%
(\phi_{N}-\sqrt{2}\phi_{S})^{2}}}\;,\quad Z_{\eta_{S}}=\frac{m_{f_{1S}}}%
{\sqrt{m_{f_{1S}}^{2}-2g_{1}^{2}\phi_{S}^{2}}}\; \label{z2}%
\end{equation}
are the wave-function renormalization constants and%
\begin{equation}
w_{f_{1N}}=w_{a_{1}}=\frac{g_{1}\phi_{N}}{m_{a_{1}}^{2}}\;,\quad w_{f_{1S}%
}=\frac{\sqrt{2}g_{1}\phi_{S}}{m_{f_{1S}}^{2}}\text{ ,} \label{wf1}%
\end{equation}%
\begin{equation}
w_{K^{\star}}=\frac{ig_{1}(\phi_{N}-\sqrt{2}\phi_{S})}{2m_{K^{\star}}^{2}%
}\;,\quad w_{K_{1}}=\frac{g_{1}(\phi_{N}+\sqrt{2}\phi_{S})}{2m_{K_{1}}^{2}%
}\;\text{.} \label{wk}%
\end{equation}
Explicit breaking of chiral symmetry is incorporated by the following constant
matrices,
\begin{equation}
H=H_{0}T_{0}+H_{8}T_{8}=\left(
\begin{array}
[c]{ccc}%
\frac{h_{0N}}{2} & 0 & 0\\
0 & \frac{h_{0N}}{2} & 0\\
0 & 0 & \frac{h_{0S}}{\sqrt{2}}%
\end{array}
\right)  \text{ ,} \label{esb1}%
\end{equation}%
\begin{equation}
E=E_{0}T_{0}+E_{8}T_{8}=\left(
\begin{array}
[c]{ccc}%
\frac{\tilde{\epsilon}_{N}}{2} & 0 & 0\\
0 & \frac{\tilde{\epsilon}_{N}}{2} & 0\\
0 & 0 & \frac{\tilde{\epsilon}_{S}}{\sqrt{2}}%
\end{array}
\right)  \equiv\left(
\begin{array}
[c]{ccc}%
\epsilon_{N} & 0 & 0\\
0 & \epsilon_{N} & 0\\
0 & 0 & \epsilon_{S}%
\end{array}
\right)  \text{ ,} \label{esb2}%
\end{equation}%
\begin{equation}
\Delta=\Delta_{0}T_{0}+\Delta_{8}T_{8}=\left(
\begin{array}
[c]{ccc}%
\frac{\tilde{\delta}_{N}}{2} & 0 & 0\\
0 & \frac{\tilde{\delta}_{N}}{2} & 0\\
0 & 0 & \frac{\tilde{\delta}_{S}}{\sqrt{2}}%
\end{array}
\right)  \equiv\left(
\begin{array}
[c]{ccc}%
\delta_{N} & 0 & 0\\
0 & \delta_{N} & 0\\
0 & 0 & \delta_{S}%
\end{array}
\right)  \text{ ,} \label{esb3}%
\end{equation}
where the terms in Eqs.\ (\ref{esb2}) and (\ref{esb3}) are next-to-leading
corrections in the current quark masses.

\section{Decay Widths}

\label{AppB}

In this work we compute two-body decays using the well-known formula%
\begin{equation}
\Gamma_{A\rightarrow BC}=s_{f}\mathcal{I}\frac{k_{f}}{8\pi m_{A}^{2}%
}\left\vert -i\mathcal{A}_{A\rightarrow BC}\right\vert ^{2}\text{ },
\label{tbd}%
\end{equation}
where%
\begin{equation}
k_{f}=\frac{1}{2m_{A}}\, \sqrt{m_{A}^{4}+(m_{B}^{2}-m_{C}^{2})^{2}-2m_{A}%
^{2}(m_{B}^{2}+m_{C}^{2})}\, \theta(m_{A}-m_{B}-m_{C})
\end{equation}
is the modulus of the three-momentum of one of the outgoing particles (the
moduli of the momenta are equal in the rest frame of the decaying particle)
and $\mathcal{A}_{A\rightarrow BC}$ is the decay amplitude. The symmetry
factor $s_{f}$ avoids double counting of identical Feynman diagrams and
$\mathcal{I}$ is the isospin factor which considers all subchannels of a
particular decay channel. The $\theta$ function encodes the decay threshold.

All relevant expressions for the decay processes studied in this work are
extracted from the Lagrangian (\ref{elsm}) and are presented in the following.

\subsection{Decays of the scalar-isoscalar fields into $\pi\pi$}

Following the general formula (\ref{tbd}) we obtain for the decay widths of
the scalar-isoscalar resonances into $\pi\pi$%
\begin{equation}
\Gamma_{f_{0}\rightarrow\pi\pi}=6\frac{\sqrt{\frac{m_{f_{0}}^{2}}{4}-m_{\pi
}^{2}}}{8\pi m_{f_{0}}^{2}}\left\vert -i\mathcal{A}_{f_{0}\rightarrow\pi\pi
}(m_{f_{0}})\right\vert ^{2}\text{ },
\end{equation}
where $m_{f_{0}}$ is the mass of the physical $f_{0}$ resonance. The bare
amplitudes (as functions of $m_{f_{0}}$) are%
\begin{equation}
-i\mathcal{A}_{\sigma_{N}\rightarrow\pi\pi}(m_{f_{0}})=i\left(  A_{\sigma
_{N}\pi\pi}-B_{\sigma_{N}\pi\pi}\frac{m_{f_{0}}^{2}-2m_{\pi}^{2}}{2}%
-C_{\sigma_{N}\pi\pi}m_{\pi}^{2}\right)  \text{ },
\end{equation}%
\begin{equation}
-i\mathcal{A}_{\sigma_{S}\rightarrow\pi\pi}(m_{f_{0}})=i\left(  A_{\sigma
_{S}\pi\pi}-B_{\sigma_{S}\pi\pi}\frac{m_{f_{0}}^{2}-2m_{\pi}^{2}}{2}\right)
\text{ },
\end{equation}%
\begin{equation}
-i\mathcal{A}_{G\rightarrow\pi\pi}(m_{f_{0}})=i\left(  A_{G\pi\pi}-B_{G\pi\pi
}\frac{m_{f_{0}}^{2}-2m_{\pi}^{2}}{2}\right)  \text{ },
\end{equation}
with the corresponding constants%
\begin{equation}
A_{\sigma_{N}\pi\pi}=-\left(  \lambda_{1}+\frac{\lambda_{2}}{2}\right)
Z_{\pi}^{2}\phi_{N}\text{ },
\end{equation}%
\begin{equation}
B_{\sigma_{N}\pi\pi}=-2g_{1}Z_{\pi}^{2}w_{a_{1}}+(g_{1}^{2}+\frac{h_{1}%
+h_{2}-h_{3}}{2})Z_{\pi}^{2}w_{a_{1}}^{2}\phi_{N}\text{ },
\end{equation}%
\begin{equation}
C_{\sigma_{N}\pi\pi}=-g_{1}Z_{\pi}^{2}w_{a_{1}}\text{ },
\end{equation}%
\begin{equation}
A_{\sigma_{S}\pi\pi}=-\lambda_{1}Z_{\pi}^{2}\phi_{S}\text{ },
\end{equation}%
\begin{equation}
B_{\sigma_{S}\pi\pi}=\frac{h_{1}}{2}Z_{\pi}^{2}w_{a_{1}}^{2}\phi_{S},
\end{equation}%
\begin{equation}
A_{G\pi\pi}=-\frac{m_{0}^{2}}{G_{0}}Z_{\pi}^{2}\text{ },
\end{equation}%
\begin{equation}
B_{G\pi\pi}=\frac{m_{1}^{2}}{G_{0}}Z_{\pi}^{2}w_{a_{1}}^{2}\text{ }.
\end{equation}
After performing an orthogonal transformation we obtain the amplitudes for the
physical scalar-isoscalar fields $\sigma_{N}^{\prime}\equiv f_{0}%
(1370),\sigma_{S}^{\prime}\equiv f_{0}(1500)$, and $G^{\prime}\equiv
f_{0}(1710)$:
\begin{equation}
-i\mathcal{A}_{\sigma_{N}^{\prime}\rightarrow\pi\pi}(m_{\sigma_{N}^{\prime}%
})=i\left[  \mathcal{A}_{\sigma_{N}\rightarrow\pi\pi}(m_{\sigma_{N}^{\prime}%
})b_{11}+\mathcal{A}_{\sigma_{S}\rightarrow\pi\pi}(m_{\sigma_{N}^{\prime}%
})b_{12}+\mathcal{A}_{G\rightarrow\pi\pi}(m_{\sigma_{N}^{\prime}}%
)b_{13}\right]  \text{ },
\end{equation}%
\begin{equation}
-i\mathcal{A}_{\sigma_{S}^{\prime}\rightarrow\pi\pi}(m_{\sigma_{S}^{\prime}%
})=i\left[  \mathcal{A}_{\sigma_{N}\rightarrow\pi\pi}(m_{\sigma_{S}^{\prime}%
})b_{21}+\mathcal{A}_{\sigma_{S}\rightarrow\pi\pi}(m_{\sigma_{S}^{\prime}%
})b_{22}+\mathcal{A}_{G\rightarrow\pi\pi}(m_{\sigma_{S}^{\prime}}%
)b_{23}\right]  \text{ },
\end{equation}%
\begin{equation}
-i\mathcal{A}_{G^{\prime}\rightarrow\pi\pi}(m_{G^{\prime}})=i\left[
\mathcal{A}_{\sigma_{N}\rightarrow\pi\pi}(m_{G^{\prime}})b_{31}+\mathcal{A}%
_{\sigma_{S}\rightarrow\pi\pi}(m_{G^{\prime}})b_{32}+\mathcal{A}%
_{G\rightarrow\pi\pi}(m_{G^{\prime}})b_{33}\right]  \text{ },
\end{equation}
where $b_{ij}$, $i,j=1,2,3$, are the corresponding elements of the mixing
matrix $B$ from Eq.\ (\ref{mixmat}).

\subsection{Decays of the scalar-isoscalar fields into $KK$}

Following the general formula (\ref{tbd}) we obtain for the decay widths of
the scalar-isoscalar resonances into $KK$%

\begin{equation}
\Gamma_{f_{0}\rightarrow KK}=2\frac{\sqrt{\frac{m_{f_{0}}^{2}}{4}-m_{KK}^{2}}%
}{8\pi m_{f_{0}}^{2}}\left\vert -i\mathcal{A}_{f_{0}\rightarrow KK}(m_{f_{0}%
})\right\vert ^{2}\text{ },
\end{equation}
where the bare amplitudes are%
\begin{equation}
-i\mathcal{A}_{\sigma_{N}\rightarrow KK}(m_{f_{0}})=i\left[  A_{\sigma_{N}%
KK}-\left(  B_{\sigma_{N}KK}-2C_{\sigma_{N}KK}\right)  \frac{m_{f_{0}}%
^{2}-2m_{K}^{2}}{2}+2C_{\sigma_{N}KK}m_{K}^{2}\right]  \text{ },
\end{equation}%
\begin{equation}
-i\mathcal{A}_{\sigma_{S}\rightarrow KK}(m_{f_{0}})=i\left[  A_{\sigma_{S}%
KK}-\left(  B_{\sigma_{S}KK}-2C_{\sigma_{S}KK}\right)  \frac{m_{f_{0}}%
^{2}-2m_{K}^{2}}{2}+2C_{\sigma_{N}KK}m_{K}^{2}\right]  \text{ },
\end{equation}%
\begin{equation}
-i\mathcal{A}_{G\rightarrow KK}(m_{f_{0}})=i\left(  A_{GKK}-B_{GKK}%
\frac{m_{f_{0}}^{2}-2m_{K}^{2}}{2}\right)  \text{ }%
\end{equation}
and the corresponding constants read%

\begin{equation}
A_{\sigma_{N}KK}=\frac{Z_{K}^{2}}{\sqrt{2}}\left[  \lambda_{2}\left(  \phi
_{S}-\sqrt{2}\phi_{N}\right)  -2\sqrt{2}\lambda_{1}\phi_{N}\right]  \text{ },
\end{equation}%
\begin{equation}
B_{\sigma_{N}KK}=\frac{g_{1}}{2}Z_{K}^{2}w_{K_{1}}\left[  -2+g_{1}w_{K_{1}%
}\left(  \phi_{N}+\sqrt{2}\phi_{S}\right)  \right]  +\frac{Z_{K}^{2}w_{K_{1}%
}^{2}}{2}\left[  \left(  2h_{1}+h_{2}\right)  \phi_{N}-\sqrt{2}h_{3}\phi
_{S}\right]  \text{ },
\end{equation}%
\begin{equation}
C_{\sigma_{N}KK}=\frac{g_{1}}{2}Z_{K}^{2}w_{K_{1}}\text{ },
\end{equation}%
\begin{equation}
A_{\sigma_{S}KK}=\frac{Z_{K}^{2}}{\sqrt{2}}\left[  \lambda_{2}\left(  \phi
_{N}-2\sqrt{2}\phi_{S}\right)  -2\sqrt{2}\lambda_{1}\phi_{S}\right]  \text{ },
\end{equation}%
\begin{equation}
B_{\sigma_{S}KK}=\frac{\sqrt{2}g_{1}}{2}Z_{K}^{2}w_{K_{1}}\left[
-2+g_{1}w_{K_{1}}\left(  \phi_{N}+\sqrt{2}\phi_{S}\right)  \right]
+\frac{Z_{K}^{2}w_{K_{1}}^{2}}{\sqrt{2}}\left[  \sqrt{2}\left(  h_{1}%
+h_{2}\right)  \phi_{S}-h_{3}\phi_{N}\right]  \text{ },
\end{equation}%
\begin{equation}
C_{\sigma_{S}KK}=\frac{\sqrt{2}g_{1}}{2}Z_{K}^{2}w_{K_{1}}\text{ },
\end{equation}

\begin{equation}
A_{GKK}=-\frac{2m_{0}^{2}}{G_{0}}Z_{K}^{2}\text{ },
\end{equation}

\begin{equation}
B_{GKK}=\frac{2m_{1}^{2}}{G_{0}}Z_{K}^{2}w_{K_{1}}^{2}\text{ }.
\end{equation}
After performing an orthogonal transformation we obtain the amplitudes for the
physical scalar-isoscalar fields%
\begin{equation}
-i\mathcal{A}_{\sigma_{N}^{\prime}\rightarrow KK}(m_{\sigma_{N}^{\prime}%
})=i\left[  \mathcal{A}_{\sigma_{N}\rightarrow KK}(m_{\sigma_{N}^{\prime}%
})b_{11}+\mathcal{A}_{\sigma_{S}\rightarrow KK}(m_{\sigma_{N}^{\prime}}%
)b_{12}+\mathcal{A}_{G\rightarrow KK}(m_{\sigma_{N}^{\prime}})b_{13}\right]  ,
\end{equation}%
\begin{equation}
-i\mathcal{A}_{\sigma_{S}^{\prime}\rightarrow KK}(m_{\sigma_{S}^{\prime}%
})=i\left[  \mathcal{A}_{\sigma_{N}\rightarrow KK}(m_{\sigma_{S}^{\prime}%
})b_{21}+\mathcal{A}_{\sigma_{S}\rightarrow KK}(m_{\sigma_{S}^{\prime}}%
)b_{22}+\mathcal{A}_{G\rightarrow KK}(m_{\sigma_{S}^{\prime}})b_{23}\right]
\text{ },
\end{equation}%
\begin{equation}
-i\mathcal{A}_{G^{\prime}\rightarrow KK}(m_{G^{\prime}})=i\left[
\mathcal{A}_{\sigma_{N}\rightarrow KK}(m_{G^{\prime}})b_{31}+\mathcal{A}%
_{\sigma_{S}\rightarrow KK}(m_{G^{\prime}})b_{32}+\mathcal{A}_{G\rightarrow
KK}(m_{G^{\prime}})b_{33}\right]  \text{ },
\end{equation}
which we assign to the physical resonances as follows: $\sigma_{N}^{\prime
}\equiv f_{0}(1370),\sigma_{S}^{\prime}\equiv f_{0}(1500),$ and $G\equiv
f_{0}(1710).$

\subsection{Decays of the scalar-isoscalar fields into $\eta\eta$}

Following the general formula (\ref{tbd}) we obtain for the decay widths of
the scalar-isoscalar resonances into $\eta\eta$%

\begin{equation}
\Gamma_{f_{0}\rightarrow\eta\eta}=2\frac{\sqrt{\frac{m_{f_{0}}^{2}}{4}%
-m_{\eta}^{2}}}{8\pi m_{f_{0}}^{2}}\left\vert -i\mathcal{A}_{f_{0}%
\rightarrow\eta\eta}(m_{f_{0}})\right\vert ^{2}\text{ },
\end{equation}
where the bare amplitudes are%

\begin{equation}
-i\mathcal{A}_{\sigma_{N}\rightarrow\eta\eta}(m_{f_{0}})=i\left(
A_{\sigma_{N}\eta\eta}-B_{\sigma_{N}\eta\eta}\frac{m_{f_{0}}^{2}-2m_{\eta}%
^{2}}{2}+C_{\sigma_{N}\eta\eta}\frac{m_{f_{0}}^{2}}{2}\right)  \text{ },
\end{equation}

\begin{equation}
-i\mathcal{A}_{\sigma_{S}\rightarrow\eta\eta}(m_{f_{0}})=i\left(
A_{\sigma_{S}\eta\eta}-B_{\sigma_{S}\eta\eta}\frac{m_{f_{0}}^{2}-2m_{\eta}%
^{2}}{2}+C_{\sigma_{S}\eta\eta}\frac{m_{f_{0}}^{2}}{2}\right)  \text{ },
\end{equation}%
\begin{equation}
-i\mathcal{A}_{G\rightarrow\eta\eta}(m_{f_{0}})=i\left[  \left(  A_{G\eta
_{N}\eta_{N}}+B_{G\eta_{N}\eta_{N}}\frac{m_{f_{0}}^{2}-2m_{\eta}^{2}}%
{2}\right)  \cos\varphi_{\eta}+\left(  A_{G\eta_{S}\eta_{S}}+B_{G\eta_{S}%
\eta_{S}}\frac{m_{f_{0}}^{2}-2m_{\eta}^{2}}{2}\right)  \sin\varphi_{\eta
}\right]  \text{ }%
\end{equation}

and the corresponding constants read%
\begin{equation}
A_{\sigma_{N}\eta\eta}=-Z_{\pi}^{2}\phi_{N}\left(  \lambda_{1}+\frac
{\lambda_{2}}{2}+c_{1}\phi_{S}^{2}\right)  \cos^{2}\varphi_{\eta}-Z_{\eta_{S}%
}^{2}\phi_{N}\left(  \lambda_{1}+\frac{c_{1}}{2}\phi_{N}^{2}\right)  \sin
^{2}\varphi_{\eta}-\frac{3}{4}c_{1}Z_{\pi}Z_{\eta_{S}}\phi_{N}^{2}\phi_{S}%
\sin(2\varphi_{\eta})\text{ },
\end{equation}%
\begin{equation}
B_{\sigma_{N}\eta\eta}=-\frac{Z_{\pi}^{2}w_{a_{1}}^{2}}{\phi_{N}}(m_{1}%
^{2}+\frac{h_{1}}{2}\phi_{S}^{2}+2\delta_{N})\cos^{2}\varphi_{\eta}%
+\frac{h_{1}}{2}Z_{\eta_{S}}^{2}w_{f_{1S}}^{2}\phi_{N}\sin^{2}\varphi_{\eta
}\text{ },
\end{equation}%
\begin{equation}
C_{\sigma_{N}\eta\eta}=g_{1}Z_{\pi}^{2}w_{a_{1}}\cos^{2}\varphi_{\eta}\text{
},
\end{equation}%
\begin{equation}
A_{\sigma_{S}\eta\eta}=-Z_{\eta_{S}}^{2}\phi_{S}\left(  \lambda_{1}%
+\lambda_{2}\right)  \sin^{2}\varphi_{\eta}-Z_{\pi}^{2}\phi_{S}\left(
\lambda_{1}+c_{1}\phi_{N}^{2}\right)  \cos^{2}\varphi_{\eta}-\frac{1}{4}%
c_{1}Z_{\pi}Z_{\eta_{S}}\phi_{N}^{3}\sin(2\varphi_{\eta})\text{ },
\end{equation}%
\begin{equation}
B_{\sigma_{S}\eta\eta}=-\frac{Z_{\eta_{S}}^{2}w_{f_{1S}}^{2}}{\phi_{S}}%
(m_{1}^{2}+\frac{h_{1}}{2}\phi_{N}^{2}+2\delta_{S})\sin^{2}\varphi_{\eta
}+\frac{h_{1}}{2}Z_{\pi}^{2}w_{a_{1}}^{2}\phi_{S}\cos^{2}\varphi_{\eta}\text{
},
\end{equation}%
\begin{equation}
C_{\sigma_{S}\eta\eta}=\sqrt{2}g_{1}Z_{\eta_{S}}^{2}w_{f_{1S}}\sin^{2}%
\varphi_{\eta}\text{ },
\end{equation}

\begin{equation}
A_{G\eta_{N}\eta_{N}}=-\frac{m_{0}^{2}}{G_{0}}Z_{\pi}^{2}\text{ },
\end{equation}

\begin{equation}
B_{G\eta_{N}\eta_{N}}=-\frac{m_{1}^{2}}{2G_{0}}Z_{\pi}^{2}w_{f_{1N}}^{2}\text{
},
\end{equation}%
\begin{equation}
A_{G\eta_{S}\eta_{S}}=-\frac{m_{0}^{2}}{G_{0}}Z_{\eta_{S}}^{2}\text{ },
\end{equation}

\begin{equation}
B_{G\eta_{S}\eta_{S}}=-\frac{m_{1}^{2}}{2G_{0}}Z_{\eta_{S}}^{2}w_{f_{1S}}%
^{2}\text{ }.
\end{equation}

After performing an orthogonal transformation we obtain the amplitudes for the
physical scalar-isoscalar fields%

\begin{equation}
-i\mathcal{A}_{\sigma_{N}^{\prime}\rightarrow\eta\eta}(m_{\sigma_{N}^{\prime}%
})=i\left[  \mathcal{A}_{\sigma_{N}\rightarrow\eta\eta}(m_{\sigma_{N}^{\prime
}})b_{11}+\mathcal{A}_{\sigma_{S}\rightarrow\eta\eta}(m_{\sigma_{N}^{\prime}%
})b_{12}+\mathcal{A}_{G\rightarrow\eta\eta}(m_{\sigma_{N}^{\prime}}%
)b_{13}\right]  \text{ },
\end{equation}

\begin{equation}
-i\mathcal{A}_{\sigma_{S}^{\prime}\rightarrow\eta\eta}(m_{\sigma_{S}^{\prime}%
})=i\left[  \mathcal{A}_{\sigma_{N}\rightarrow\eta\eta}(m_{\sigma_{S}^{\prime
}})b_{21}+\mathcal{A}_{\sigma_{S}\rightarrow\eta\eta}(m_{\sigma_{S}^{\prime}%
})b_{22}+\mathcal{A}_{G\rightarrow\eta\eta}(m_{\sigma_{S}^{\prime}}%
)b_{23}\right]  \text{ },
\end{equation}

\begin{equation}
-i\mathcal{A}_{G^{\prime}\rightarrow\eta\eta}(m_{G^{\prime}})=i\left[
\mathcal{A}_{\sigma_{N}\rightarrow\eta\eta}(m_{G^{\prime}})b_{31}%
+\mathcal{A}_{\sigma_{S}\rightarrow\eta\eta}(m_{G^{\prime}})b_{32}%
+\mathcal{A}_{G\rightarrow\eta\eta}(m_{G^{\prime}})b_{33}\right]  \text{ },
\end{equation}
which we assign to the physical resonances as follows: $\sigma_{N}^{\prime
}\equiv f_{0}(1370),\sigma_{S}^{\prime}\equiv f_{0}(1500),$ and $G\equiv
f_{0}(1710).$

\subsection{Decays of the scalar-isoscalar fields into $\rho\rho
\rightarrow4\pi$}

The decay processes $f_{0}\rightarrow\rho\rho\rightarrow4\pi$ are on the
threshold, hence we use for the calculation of the decay widths the spectral
function of the $\rho$ meson
\begin{equation}
d_{\rho}(X_{m_{\rho}})=N\frac{X_{m_{\rho}}^{2}\Gamma_{\rho\rightarrow\pi\pi
}(X_{m_{\rho}})}{(X_{m_{\rho}}^{2}-m_{\rho}^{2})^{2}+X_{m_{\rho}}^{2}%
\Gamma_{\rho\rightarrow\pi\pi}^{2}(X_{m_{\rho}})}\theta(X_{m_{\rho}}-2m_{\pi
})\text{ },
\end{equation}
where $N$ is a normalization constant. Considering the polarization of the
$\rho$ mesons the general amplitude reads
\begin{equation}
\overline{\left\vert -i\mathcal{A}_{f_{0}\rightarrow\rho\rho}(m_{f_{0}%
},X_{i,m_{\rho}})\right\vert }^{2}=A_{\rho\rho}^{2}\left[  4-\frac
{X_{1,m_{\rho}}^{2}+X_{2,m_{\rho}}^{2}}{m_{\rho}^{2}}+\frac{(m_{f_{0}}%
^{2}-X_{1,m_{\rho}}^{2}-X_{2,m_{\rho}}^{2})^{2}}{4m_{\rho}^{4}}\right]  \text{
},
\end{equation}
where $i=1,2$ and $A_{\rho\rho}$ is one of the corresponding constants%
\begin{equation}
A_{\sigma_{N}\rho\rho}=\frac{\phi_{N}}{2}\left(  h_{1}+h_{2}+h_{3}\right)
\text{ },
\end{equation}%
\begin{equation}
A_{\sigma_{S}\rho\rho}=\frac{\phi_{S}}{2}h_{1}\text{ },
\end{equation}

\begin{equation}
A_{G\rho\rho}=\frac{m_{1}^{2}}{G_{0}}\text{ }.
\end{equation}
The physical amplitudes of the scalar-isoscalar fields read%

\begin{align}
&  \overline{\left\vert -i\mathcal{A}_{\sigma_{N}^{\prime}\rightarrow\rho\rho
}(m_{f_{0}},X_{i,m_{\rho}})\right\vert }^{2}\nonumber\\
&  =\left[  A_{\sigma_{N}\rho\rho}b_{11}+A_{\sigma_{S}\rho\rho}b_{12}%
+A_{G\rho\rho}b_{13}\right]  ^{2}\left[  4-\frac{X_{1,m_{\rho}}^{2}%
+X_{2,m_{\rho}}^{2}}{m_{\rho}^{2}}+\frac{(m_{f_{0}}^{2}-X_{1,m_{\rho}}%
^{2}-X_{2,m_{\rho}}^{2})^{2}}{4m_{\rho}^{4}}\right]  \text{ },
\end{align}

\begin{align}
&  \overline{\left\vert -i\mathcal{A}_{\sigma_{S}^{\prime}\rightarrow\rho\rho
}(m_{f_{0}},X_{i,m_{\rho}})\right\vert }^{2}\nonumber\\
&  =\left[  A_{\sigma_{N}\rho\rho}b_{21}+A_{\sigma_{S}\rho\rho}b_{22}%
+A_{G\rho\rho}b_{23}\right]  ^{2}\left[  4-\frac{X_{1,m_{\rho}}^{2}%
+X_{2,m_{\rho}}^{2}}{m_{\rho}^{2}}+\frac{(m_{f_{0}}^{2}-X_{1,m_{\rho}}%
^{2}-X_{2,m_{\rho}}^{2})^{2}}{4m_{\rho}^{4}}\right]  \text{ },
\end{align}

\begin{align}
&  \overline{\left\vert -i\mathcal{A}_{G^{\prime}\rightarrow\rho\rho}%
(m_{f_{0}},X_{i,m_{\rho}})\right\vert }^{2}\nonumber\\
&  =\left[  A_{\sigma_{N}\rho\rho}b_{31}+A_{\sigma_{S}\rho\rho}b_{32}%
+A_{G\rho\rho}b_{33}\right]  ^{2}\left[  4-\frac{X_{1,m_{\rho}}^{2}%
+X_{2,m_{\rho}}^{2}}{m_{\rho}^{2}}+\frac{(m_{f_{0}}^{2}-X_{1,m_{\rho}}%
^{2}-X_{2,m_{\rho}}^{2})^{2}}{4m_{\rho}^{4}}\right]  \text{ }.
\end{align}
The formula for the decays of the scalar-isoscalar fields into $\rho$ mesons
and 4$\pi$, respectively, reads%
\begin{equation}
\Gamma_{f_{0}\rightarrow\rho\rho}(m_{f_{0}},X_{i,m_{\rho}})=6\frac
{k_{f}(m_{f_{0}},X_{i,m_{\rho}})}{8\pi m_{f_{0}}^{2}}\overline{\left\vert
-i\mathcal{A}_{f_{0}\rightarrow\rho\rho}(m_{f_{0}},X_{i,m_{\rho}})\right\vert
}^{2}\theta(m_{f_{0}}-X_{1,m_{\rho}}-X_{2,m_{\rho}})\text{ },
\end{equation}

\begin{equation}
\Gamma_{f_{0}\rightarrow\rho\rho\rightarrow4\pi}=\int_{0}^{\infty}\int
_{0}^{\infty}\Gamma_{f_{0}\rightarrow\rho\rho}(m_{f_{0}},X_{i,m_{\rho}%
})d_{\rho}(X_{1,m_{\rho}})d_{\rho}(X_{2,m_{\rho}})dX_{1,m_{\rho}}%
dX_{2,m_{\rho}}\text{ }.
\end{equation}

The scalar-isoscalar fields are assigned to the physical resonances as
follows: $\sigma_{N}^{\prime}\equiv f_{0}(1370),\sigma_{S}^{\prime}\equiv
f_{0}(1500),$ and $G\equiv f_{0}(1710).$

\end{document}